\documentclass[secnumarabic,amssymb,superscriptaddress,aps,prl,reprint]{revtex4-1}

\usepackage[utf8]{inputenc}
\usepackage[T1]{fontenc}
\usepackage{multirow}
\usepackage[table]{xcolor}
\let\phi\varphi

\usepackage{graphicx}
\usepackage{titlesec}
\usepackage{color}
\usepackage{dcolumn}
\usepackage{bm}
\usepackage{ucs}
\usepackage{physics}
\titlespacing\section{0pt}{12pt plus 4pt minus 2pt}{0pt plus 2pt minus 2pt}
\usepackage{mathtools}
\begin{document}

\title{Magnetic texture modulated superconductivity in superconductor/ferromagnet shells of semiconductor nanowires}

\author{Nabhanila Nandi}
\email{nandi@stanford.edu}
\affiliation{Department of Physics, Stanford University, Stanford, California 94305, USA}
\affiliation{Stanford Institute for Materials and Energy Sciences, SLAC National Accelerator Laboratory, 2575 Sand Hill Road, Menlo Park, CA 94025, USA}
\author{Juan Carlos Estrada Salda\~{n}a}
\affiliation{Niels Bohr Institute, University of Copenhagen, 2100 Copenhagen, Denmark}
\author{Alexandros Vekris}
\affiliation{Niels Bohr Institute, University of Copenhagen, 2100 Copenhagen, Denmark}
\author{Michelle Turley}
\affiliation{Niels Bohr Institute, University of Copenhagen, 2100 Copenhagen, Denmark}
\author{Irene P. Zhang}
\affiliation{Stanford Institute for Materials and Energy Sciences, SLAC National Accelerator Laboratory, 2575 Sand Hill Road, Menlo Park, CA 94025, USA}
\affiliation{Department of Applied Physics, Stanford University, Stanford, California 94305, USA}
\author{Yu Liu}
\affiliation{Niels Bohr Institute, University of Copenhagen, 2100 Copenhagen, Denmark}
\author{Mario Castro}
\affiliation{Departamento de F\'isica, FCFM, Universidad de Chile, Santiago, 8370448, Chile}
\author{Martin Bjergfelt}
\affiliation{Niels Bohr Institute, University of Copenhagen, 2100 Copenhagen, Denmark}
\author{Sabbir A. Khan}
\affiliation{Niels Bohr Institute, University of Copenhagen, 2100 Copenhagen, Denmark}
\affiliation{Danish Fundamental Metrology, Kogle Alle 5, 2970 H{\o}rsholm, Denmark}
\author{Sebasti\'an Allende}
\affiliation{Departamento de Física, CEDENNA, Universidad de Santiago de Chile, Av. Ecuador 3493, Santiago 9170124, Chile}
\author{Peter Krogstrup}
\affiliation{Niels Bohr Institute, University of Copenhagen, 2100 Copenhagen, Denmark}
\author{Kathryn Ann Moler}
\affiliation{Department of Physics, Stanford University, Stanford, California 94305, USA}
\affiliation{Stanford Institute for Materials and Energy Sciences, SLAC National Accelerator Laboratory, 2575 Sand Hill Road, Menlo Park, CA 94025, USA}
\affiliation{Department of Applied Physics, Stanford University, Stanford, California 94305, USA}
\author{Kasper Grove-Rasmussen}
\affiliation{Niels Bohr Institute, University of Copenhagen, 2100 Copenhagen, Denmark}
\author{Jesper Nyg{\aa}rd}
\email{nygard@nbi.ku.dk}
\affiliation{Niels Bohr Institute, University of Copenhagen, 2100 Copenhagen, Denmark}

\begin{abstract}

In a one-dimensional ferromagnet-superconductor nanowire, magnetism can suppress superconductivity except where the Zeeman field is suppressed, for example domain wall superconductivity (DWS) near magnetic domain walls or multi-domain-averaged superconductivity (MDAS) in multi-domain states where the net magnetization over the coherence length averages to nearly zero. Here we study full-shell InAs/EuS/Al nanowires using scanning SQUID magnetometry and transport, and find superconductivity in the Al shell only when the EuS is in a multi-domain state, consistent with both DWS and MDAS, and absent in the saturated single-domain state. Scanning SQUID magnetometry further shows that the EuS magnetic texture is position dependent and reconfigurable by small changes in external magnetic field, including moving a well-defined domain wall at $\approx 5.5~\mu\mathrm{m}/\mathrm{mT}$ with sub-mT fields, implying that any associated localized superconducting region would likewise be movable. Such magnetic texture controlled superconductivity along a nanowire may be useful for topological qubits, Andreev spin qubits, superconducting logic, and memory devices.

\end{abstract}

\flushbottom
\maketitle

\thispagestyle{empty}

\begin{figure} [h]
\centering
\includegraphics[width=0.85\linewidth]{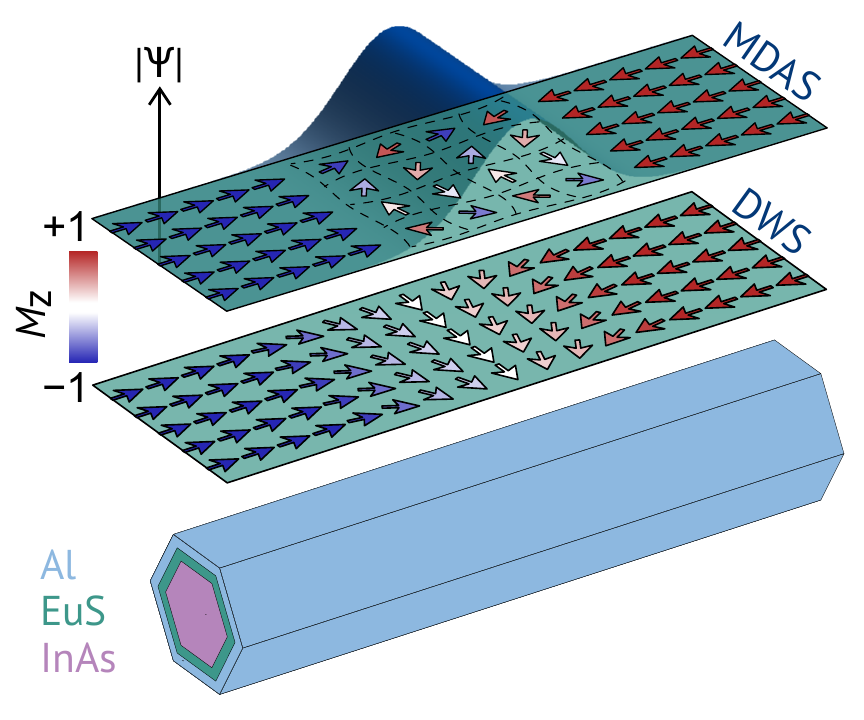}
\caption{\textbf{Domain wall superconductivity in a full-shell InAs/EuS/Al nanowire.} 
Schematic of a hexagonal nanowire with an InAs core, EuS ferromagnetic layer, and outer Al shell. The two offset planes show two plausible EuS magnetic textures in which superconductivity can survive in the Al, with the color scale indicating the EuS magnetization component along the nanowire axis, $M_z$. The lower plane has two oppositely magnetized domains ($M_z = \pm 1$) separated by a single magnetic domain wall (MDW), while the upper plane illustrates a multi-domain EuS state with many small $+M_z$ and $-M_z$ domains. A blue surface plot overlaid on the upper plane represents the superconducting order-parameter magnitude $|\Psi|$ in the Al, which is finite only where the Zeeman field from the EuS vanishes or averages to near zero over the Al coherence length, either locally at an MDW or in a multi-domain-averaged superconducting (MDAS) region, where many small EuS domains produce a near-zero net magnetization and is suppressed in regions of nearly uniform magnetization.}
\label{Fig1}
\end{figure}

Hybrid semiconductor nanowires with epitaxial superconducting shells are a well-established, versatile platform for quantum devices and unconventional superconducting phases~\cite{ Prada2020Oct, Aguado2020Dec, Badawy2024Mar,pitavidal2025novelqubitshybridsemiconductorsuperconductor}.
In particular, epitaxially grown InAs/Al heterostructure nanowires combine strong spin-orbit coupling and superconducting proximity effects in a geometry compatible with gate tunability and hybrid circuit integration. 
These materials have been used in the search for Majorana bound states in one-dimensional and hybrid quantum dot geometries \cite{Lutchyn2018May, Prada2020Oct,Dvir2023Feb}, and enabled the exploration of Andreev spin qubits, in which spin couples to supercurrent via spin-orbit interaction \cite{Tosi2019Jan, Hayes2021SpinQubitNW, Pita-Vidal2023SpinQubitNW}. Superconducting hybrid nanowires have also been used in demonstrations of gate-tunable qubits, rectifiers, switches and microwave amplifiers \cite{Aguado2020Dec, Szombati2016May, Scherubl2025Aug, Splitthoff2024Jan}.
The ability to engineer and manipulate superconducting states in such nanowires has the potential to unlock new opportunities for superconducting logic, memory, and quantum information applications.

Adding a ferromagnetic layer to the heterostructure platform provides both a stray magnetic field and a ferromagnetic proximity effect, thereby creating a Zeeman shift that is theoretically useful for creating exotic states. 
These Zeeman shifts can also suppress superconductivity. 
Because Cooper pairs are spatially extended over the coherence length $\xi$, a superconductor experiences Zeeman field spatially averaged over this length scale. 
One can therefore expect superconductivity to survive wherever the averaged Zeeman field is suppressed: near magnetic domain walls (MDWs), giving rise to domain wall superconductivity (DWS), and in a multi-domain state where individual domains are smaller than $\xi$, so that their net magnetization averages to nearly zero over $\xi$, a scenario we refer to as multi-domain-averaged superconductivity (MDAS).
In the quasi-1D limit, shape anisotropy favors axial magnetic domains and well-defined MDWs. 
Such domain textures can host highly localized superconducting regions that can be repositioned along the nanowire by changing the external magnetic field or other control knobs.
Spatially confined DWS has been observed by transport and imaging in quasi-2D heterostructures~\cite{Yang2004Nov,Iavarone2014Aug}.

Epitaxial EuS/Al bilayer grown on InAs nanowires~\cite{Liu2020Jan,Vaitiekenas2022Jan,Razmadze2023Feb} provides a natural platform for tunable superconductivity: the superconductor Al shell proximizes the InAs~\cite{Mayer2019Jan,Bubis2017Aug}, while the ferromagnet EuS imparts a Zeeman field to both InAs and Al~\cite{Liu2020Feb,Liu2021Jul}. In this work, we focus on the direct interplay between this superconductivity and the ferromagnetism of the EuS/Al shell. Previous work shows that Al in EuS/Al nanowire heterostructures experiences a large effective Zeeman field, comparable to the exchange field $H_\mathrm{exch}$ inferred in planar EuS/Al films ($0.5$--$4.5$~T), which strongly suppresses superconductivity~\cite{Xiong2011Jun,Strambini2017Oct,Moodera1988Aug,Hao1991Sep,Geng2023Oct,maiani2024percolative}.
Spatial variations of the magnetization texture can locally average this field over the Al coherence length $\xi_\mathrm{Al} \approx 200$~nm, allowing superconductivity to reemerge either as DWS associated with nanoscale to micron-scale MDWs~\cite{Houzet2006Dec} or more generally via Zeeman-field averaging over multiple nanoscale domains.
Superconductivity observed near the coercive field in partially EuS/Al covered InAs nanowires may be due to this mechanism~\cite{Vaitiekenas2020Apr, Razmadze2023Feb}.

In this work, we study InAs nanowires fully coated on all six facets with a EuS/Al bilayer.  
Using scanning SQUID magnetometry together with transport, we demonstrate the existence of magnetic texture modulated superconductivity in these heterostructure nanowires, correlate where superconductivity appears in Al with the underlying EuS magnetic domain texture, and explore its dependence on field, current, temperature, and time. 
By tuning the vector magnetic field, we demonstrate controlled manipulation of the domain texture and, in turn, the superconducting phase.
Notably, in the vicinity of the coercive field, a single MDW forms, and with only sub-mT external field this MDW can be moved along the nanowire with micron-scale precision, so that any DWS tied to it would be correspondingly movable.

\begin{figure}[tp]
\centering
\includegraphics[width=\linewidth]{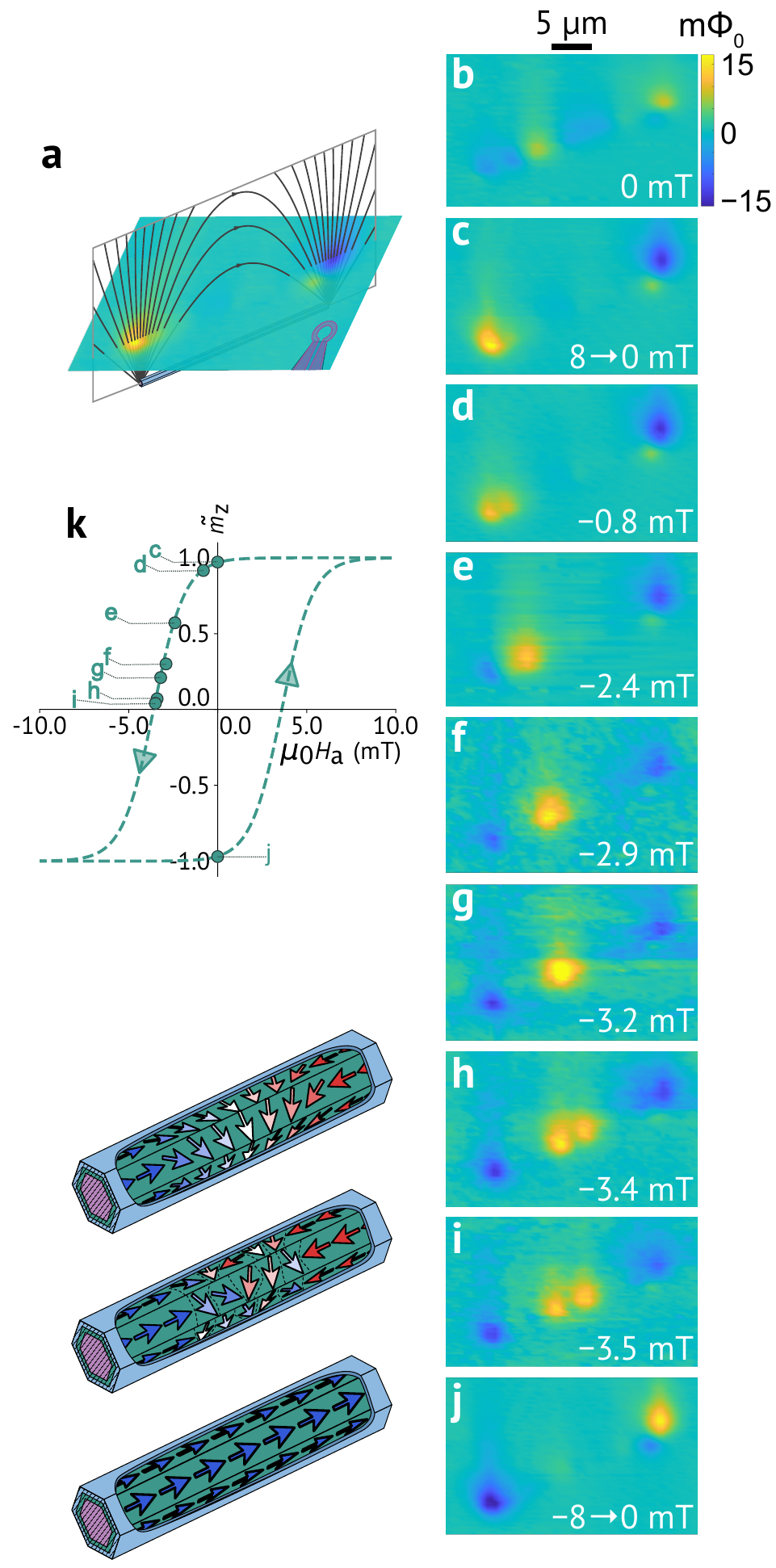}
\caption{\textbf{Magnetic texture modulated superconductivity and SQUID magnetometry of domain formation in the EuS shell of an InAs/EuS/Al full-shell nanowire NW1.}
\textbf{a} Schematic of a nanowire, shown as a thin blue cylinder, with the expected field lines in a fully magnetized, single-domain configuration. The scanning SQUID maps the out-of-plane magnetic flux through its pick-up loop, and the measured flux appears as two lobes of opposite sign at the wire ends.
\textbf{b}--\textbf{j} Scanning SQUID magnetometry images as the field along the nanowire is swept from $0 \to +8$~mT, $+8 \to -8$~mT, and then $-8 \to 0$~mT.
Next to \textbf{h}, \textbf{i}, and \textbf{j} we show conceptual sketches of plausible magnetization configurations in the EuS shell. The sketches next to \textbf{h} and \textbf{i} illustrate two alternative configurations that could each account for the similar SQUID images in both panels.
These schematics are not to scale and represent only the central portion of the nanowire.
\textbf{k} Sketch of the corresponding hysteresis loop, showing the nanowire’s normalized axial magnetic moment $\tilde{m}_\text{z}$ versus the applied axial field $H_\mathrm{a}$. The circles are computed from the SQUID images, as described in the End Matter, and the dashed curve is a guide to the eye for the conceptual $\tilde{m}_\text{z}$--$H_\text{a}$ curve.
All scans are taken at 4.2~K.}

\label{Fig2}
\end{figure}

Epitaxial layers of $3-10$ nm EuS and $4-10$ nm Al coat all six facets of InAs nanowires with typical lengths of $10-20$ $\mu$m and diameters of $100-250$~nm, a schematic is shown in Fig.~\ref{Fig1}. 
Because the Al film is thin and has an epitaxial interface with the EuS, we expect a strong $H_\mathrm{exch}$ throughout the Al.
The main text focuses on measurements as a function of field, while those as a function of current, temperature, and time are presented in the Supplemental Material (SM).
We first present the results of the magnetometry and transport measurements, then discuss how the observations frame DWS and MDAS as two plausible mechanisms for the superconducting phase.

Scanning SQUID has proven useful for studying domain formation in InAs nanowires with facets partially covered with EuS \cite{Liu2020Jan}, but no imaging studies of full-shell nanowires in field have been reported to our knowledge. We image the stray field of InAs/EuS/Al nanowires using SQUID magnetometry, where the SQUID measures the flux through its pickup loop, i.e., the sample's $z$-field convolved with the loop geometry. A single domain in a quasi-1D system will have field lines leaving and entering each end of the domain appearing as a pair of resolution-limited ($\approx 1~\mu$m, see SM Fig.~S8) positive and negative lobes at the wire ends (Fig.~\ref{Fig2}a).

We image two bare, non-contacted nanowires. Figs.~\ref{Fig2}b--j show magnetometry images of NW1 as an axial field $H_\text{a}$ is swept through a hysteresis loop, 0~mT to $+8$~mT, to $-8$~mT, and back to 0~mT. In the zero-field-cooled state (Fig.~\ref{Fig2}b), multiple positive and negative lobes along the wire indicate a multi-domain configuration with a combination of micron-scale and nanoscale domains.
Sweeping to $+8$~mT trains the nanowire into a single domain, which persists as the field is swept back to 0~mT, indicated by a single positive and negative lobe at the nanowire's ends (Fig.~\ref{Fig2}c).
Sweeping into negative fields, a second domain nucleates at the left end of the nanowire around the lower coercive field \(H_{\mathrm{cl}}=-0.8~\mathrm{mT}\) (Fig.~\ref{Fig2}d). 
By \(-2.4~\mathrm{mT}\) (Fig.~\ref{Fig2}e), a negative lobe at the left end and a positive lobe a short distance along the nanowire indicate the clear formation of a second domain, separated by an unresolved MDW.
Increasing the field further, the MDW starts to move along the nanowire at a rate of $\approx5.5~\mu$m/mT (Figs.~\ref{Fig2}e--g). 
Near the nominal coercive field $H_{\text{c}} \approx -3.5$~mT, the two domains are comparable in size and a few-micron region of near-zero signal appears between them (Fig.~\ref{Fig2}h, j), consistent with either an extended MDW or many nanoscale domains.
In the two nanowires of different lengths that we imaged, this near-zero signal extends over the same length, approximately $3.7~\mu\mathrm{m}$ (for NW2, see Fig.~\ref{Fig4}c). 
At this nominal coercive field, $H_{\text{c}}=-3.5$~mT, the magnetic moment of the nanowire becomes zero.

From this point up to saturation, we could not image NW1 because of experimental constrains. 
To access the corresponding field range, we instead look at the image from NW2 in Fig.~\ref{Fig4}d. 
As the field is swept further, the portion of the original domain that remained at $H_{\text{c}}$ evolves into multiple smaller domains.
We cannot unambiguously reconstruct the detailed domain pattern, but we infer that the measured signal reflects a mixture of nano to sub-micron scale domains, yielding a spatially averaged contrast that still points to a predominantly reversed polarity to the initial domain.
Increasing the field above the saturation field $H_{\text{sat}}$ magnetizes the nanowire back into a single domain, but in the opposite direction, that again sustains as the field is swept back to zero (Fig.~\ref{Fig2}j).
We calculate the normalized axial magnetic moment $\tilde{m}_\text{z}$ as a function of applied field $H_\text{a}$ from the magnetometry images in Fig.~\ref{Fig2}b--j (details in End Matter (EM)) and plot along with a conceptual hysteresis loop in Fig.~\ref{Fig2}k.

\begin{figure*}[t!]
\centering
\includegraphics[width=0.75\linewidth]{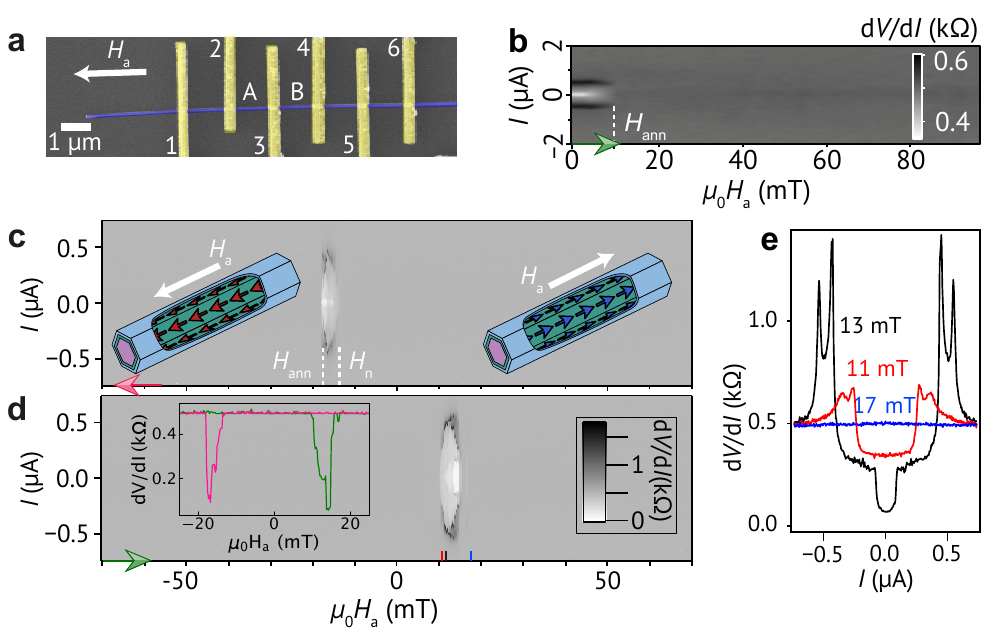}
\caption{\textbf{Bulk transport measurements on an InAs/EuS/Al full-shell nanowire while sweeping a magnetic field $H_\text{a}$ parallel to the wire.} 
\textbf{a} False-color scanning electron micrograph of a nanowire with transport contacts (1--6).
\textbf{b} Four-terminal differential resistance $\mathrm{d}V/\mathrm{d}I(I, H_\text{a})$ of segment A after zero-field cooling, for $H_\text{a}$ swept from 0 $\to$ +100 mT. A low-resistance region appears near zero field and vanishes at an annihilation field $H_\text{ann}$, which we interpret as a superconducting phase. 
\textbf{c,d} $\mathrm{d}V/\mathrm{d}I(I, H_\text{a})$ for field sweeps (c) +70 mT $\to$ $-$70 mT and (d) $-$70 mT $\to$ +70 mT. Superconductivity is confined to a narrow field window, nucleating at $H_\text{n}$ and disappearing at $H_\text{ann}$, which we associate with the coercive field. Green and pink arrows indicate sweep directions. Insets in (c) show plausible single-domain magnetization configurations in the EuS shell as $H_\text{a}$ is swept to saturation in either direction. The inset in (d) shows zero-bias $\mathrm{d}V/\mathrm{d}I$ as a function of $H_\text{a}$ for both field-sweep directions.
\textbf{e} $\mathrm{d}V/\mathrm{d}I(I)$ line cuts from (d) at $H_\text{a} = 11$, 13, and 17 mT. All data are taken at 30 mK.
}
\label{Fig3}
\end{figure*}

Superconductivity in the Al is not detected in our 4.2 K magnetometry images because its critical temperature is $\lesssim 1$ K, and even at lower temperatures the Meissner response of a quasi-1D shell would likely be too weak to resolve. 
We therefore probe the superconductivity in the Al shell via four-probe differential resistance $\mathrm{d}V/\mathrm{d}I$ measurements at 30 mK. 
In comparing magnetometry and transport, we note that the magnetometry was performed at 4.2~K; while we would expect qualitatively similar magnetic behavior at lower temperatures, it would likely occur at a higher field.
In the main text, we present data from two 700 nm segments, A and B, of a multi-contact transport device (Fig.~\ref{Fig3}a). Fig.~\ref{Fig3} shows data from segment A, with current sourced between contacts 1 and 6 and voltage measured between contacts 2 and 3.
We measure $\mathrm{d}V/\mathrm{d}I$ vs. dc bias current, $I$, for a series of field sweeps. 
Throughout most of each sweep (Figs.~\ref{Fig3}b--e), $\mathrm{d}V/\mathrm{d}I$ is independent of $H_\text{a}$ and $I$, identifying the normal-state resistance \(R_{\text{n}} = 0.5\,\mathrm{k}\Omega\). 

In the zero-field-cooled state (Fig.~\ref{Fig3}b), $\mathrm{d}V/\mathrm{d}I$ shows structure for $\lvert I\rvert \lesssim 500$~nA, with a minimum near zero-bias and peaks at $\lvert I\rvert \approx 500$~nA; this feature persists as $H_\text{a}$ is swept and disappears at the annihilation field $H_{\text{ann}} \approx10$~mT. When the field is swept from $+70$ to $-70$~mT (Fig.~\ref{Fig3}c), a similar feature reappears at $H_\text{n} = -13$~mT and vanishes at $H_{\text{ann}} = -18$~mT; on the reverse sweep (Fig.~\ref{Fig3}d) it returns at $H_\text{n} = 10$~mT and disappears at $H_{\text{ann}} = 17$~mT, showing similar but not perfectly symmetric behavior in field sweep direction. The inset to Fig.~\ref{Fig3}d summarizes this asymmetry in zero-bias $\mathrm{d}V/\mathrm{d}I$ versus $H_\text{a}$. 

Fig.~\ref{Fig3}e presents $\mathrm{d}V/\mathrm{d}I$ vs $I$ linecuts at three representative fields from Fig.~\ref{Fig3}d. At 11 mT and 13 mT, $\mathrm{d}V/\mathrm{d}I$ is suppressed at low bias with two symmetric finite-bias peaks at higher bias, signifying the superconducting state. The trace at 13~mT exhibits a lower resistance minimum than at 11~mT. Additionally, at 13~mT the resistance drops abruptly at a bias current of $I=93$~nA. The 17~mT trace shows flat resistance, consistent with the normal state.

Across six segments in three transport devices (Figs.~\ref{Fig3}, \ref{Fig4}n, and Fig.~\ref{Fig_em_1}), we observe the same phenomenology: reduced resistance at low bias, one or two intermediate-bias peaks, and sharp resistance jumps at specific currents and fields. We interpret this as superconductivity confined to a narrow range of parameters. Comparing with SQUID magnetometry images (Fig.~\ref{Fig2}b--j), we see that the superconducting window tracks a multi-domain state of the EuS shell, appearing after zero-field cooling and within the coercive field region, and disappearing once the EuS magnetization saturates, where the large exchange field suppresses superconductivity. The corresponding field range is consistent with estimates and previous measurements of $H_\mathrm{c}$ in EuS-coated nanowires and EuS/Al bilayer films at mK temperatures~\cite{Vaitiekenas2020Apr,Strambini2017Oct,Xiong2011Jun}.

\begin{figure}[tp]
\centering
\includegraphics[width=\linewidth]{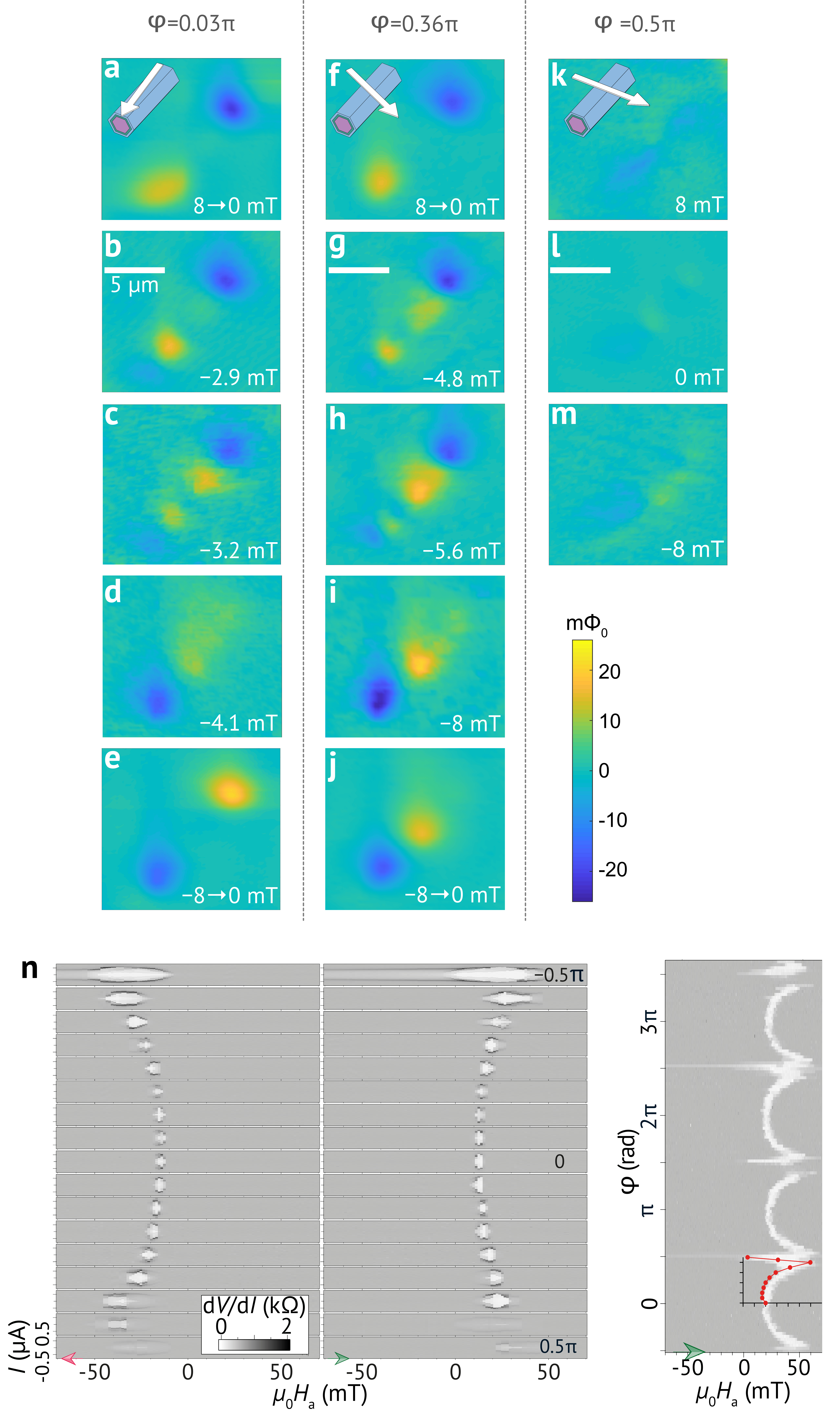}
\caption{{\bf{Field angle dependence of the superconducting phase in InAs/EuS/Al full-shell nanowires.}}
\textbf{a--m} SQUID magnetometry images of domain formation in the EuS shell in NW2 at three field angles, $\phi=0.03\pi$, $0.36\pi$, and $0.5\pi$. 
$\phi$ is the angle between the applied field and the nanowire axis.
For each $\phi$, $H_\text{a}$ is swept from $0 \to +8$~mT, $+8 \to -8$~mT, and $-8 \to 0$~mT. 
These images show how domain nucleation and propagation depend on field angle: as $\phi$ increases, the coercive field window and the saturation field shift to higher values, and the coercive-field window broadens.
These scans are taken at 4.2~K. 
\textbf{n} Field angle dependence of the differential resistance \(\mathrm{d}V/\mathrm{d}I\). 
\emph{Left} \(\mathrm{d}V/\mathrm{d}I(I,H_\text{a})\) colormaps at different \(\phi\), shown in vertically stacked panels.
Two columns correspond to opposite field sweep directions of indicated by pink and green arrows on the bottom plots.
With increasing field angle the superconducting region shifts to larger fields and persists over a larger field window.
\emph{Right} Zero-bias \(\mathrm{d}V/\mathrm{d}I(\phi,H_\text{a})\) colormap with the sweep direction indicated on the \(H_\text{a}\) axis. 
The color scale is common to all plots.
The red curve overlaid on the measured data is the simulated $H_\mathrm{c} (\phi)$ of the EuS nanotube shell.
All measurements are done on segment B of the transport device (Fig.~\ref{Fig3}a) at 30~mK.}
\label{Fig4}
\end{figure}

To further constrain the mechanism behind the superconducting phase, we rotate the applied field by an angle $\phi$ relative to the nanowire axis and image a second non-contacted nanowire NW2 as $H_\text{a}$ is swept through a full hysteresis loop, 0~mT to $+8$~mT, to $-8$~mT, and back to 0~mT. Fig.~\ref{Fig4}a--m shows images at three angles, $\phi = 0.03\pi$, $0.36\pi$, and $0.5\pi$ (more angles in Fig.~\ref{Fig_em_2}). For small and intermediate angles, $0.03\pi$ and $0.36\pi$, domain formation broadly resembles NW1 at $\phi = 0$ (Fig.~\ref{Fig2}b--j), however, at $0.36\pi$ the wire is in a single-domain at $+8$ mT, whereas at $-8$ mT one extended domain coexists with fragmented, unresolved domains, that persist as the field returns to zero. At $\phi = 0.5\pi$, the largest fields $\pm 8$ mT align multiple domains predominantly perpendicular to the nanowire, which then rotate back toward the wire axis as the field is reduced to zero. Taken together, these images show that as the field angle increases, EuS coercive-field window shifts to higher fields, broadens, and supports multi-domain states over a wider range of fields.

In transport, we likewise see this evolution reflected in the superconducting response. Fig.~\ref{Fig4}n shows the angle-dependent superconducting phase in $\mathrm{d}V/\mathrm{d}I(I,H_\text{a})$ and in the zero-bias $\mathrm{d}V/\mathrm{d}I(H_\text{a},\phi)$, measured in segment B between voltage contacts 3 and 4. As $\phi$ increases, the field interval over which the superconducting phase is observed ($H_\text{n} - H_\text{ann}$) likewise shifts to higher fields and broadens, nearly spanning the full field range around $\phi = 0.5\pi$. 
The distinctive angular profile of the superconducting phase appears in every segment across nanowires where superconductivity is observed (see also Fig.~\ref{Fig_em_1}), and is largely symmetric under a full $2\pi$ rotation.
It is in agreement with micromagnetic simulations of $H_{\text{c}} (\phi)$, further supporting the link between the coercive field of EuS and superconductivity in Al.
The simulated $H_\mathrm{c} (\phi)$ is shown as a red curve overlaid on the zero-bias $\mathrm{d}V/\mathrm{d}I$ data in Fig.~\ref{Fig4}n, calculated by tracing the hysteresis curve of the EuS shell at a given $\phi$ (details in SM). 

To understand the microscopic origin of the superconducting phase, we interpret the transport signatures in light of the domain textures revealed by scanning SQUID magnetometry across multiple nanowires, field angles, and device segments. In doing so, we ask whether the superconductivity arises from DWS near a well-defined MDW, or MDAS, in which the Al coherence length $\xi_\mathrm{Al} \approx 200$~nm spans many nanoscale domains. SQUID images in the zero-field-cooled state and coercive-field window reveal a range of compatible magnetic textures: from micron-scale domains separated by either a resolution-limited MDW or a few micron near-zero flux region (e.g., Fig.~\ref{Fig2}b,d--i, Fig.~\ref{Fig4}b,c,g,h,l), to more fragmented configurations with mixed nano to micron-scale domains with sub-micron to micron-long patches of strongly reduced or vanishing magnetization (e.g., Figs.~\ref{Fig2}b, ~\ref{Fig4}d,i,j,k,m).

However, we cannot associate a unique magnetic configuration with a given $\approx 700$~nm transport segment. Each segment can host a combination of the observed textures, which the magnetometry images show to vary along the nanowire, and most likely be locally perturbed by the contacts. Without simultaneous magnetometry and transport measurements, it is therefore difficult to establish a strict one-to-one correlation. The spatial variation is reflected in transport: adjacent segments A and B in the same device display highly dissimilar superconducting windows (Figs.~\ref{Fig3}c,d and \ref{Fig4}n), with resistance drop nearly symmetric to strongly skewed in applied magnetic field. 
Across three devices, one presented in the main text and two in EM Fig.~\ref{Fig_em_1}, the superconducting phase differs markedly between devices and is asymmetric under field-sweep reversal and $0.5\pi$ field rotations, mirroring corresponding asymmetries in the SQUID images (Fig.~\ref{Fig4}a--m, EM Fig.~\ref{Fig_em_2}). 
These asymmetries likely stem from lack of radial and bilateral symmetry in the nanowire morphology, such as tapering or fabrication defects, which set domain nucleation, pinning, and the resulting superconducting response.

Within this limitation, the size and shape of the resistance drop still provide useful clues. In the zero-field-cooled state (Fig.~\ref{Fig3}b), and after field training near $\phi \approx 0.5\pi$ followed by returning to zero field (Figs.~\ref{Fig4}n,~\ref{Fig_em_1}a), we see only a modest resistance drop, whereas in the coercive-field window the drop is much larger (Figs.~\ref{Fig3}c,d,~\ref{Fig4}n,~\ref{Fig_em_1}a). SQUID images show that zero-field states are dominated by micron-scale domains with resolution limited MDWs, while the coercive regime additionally hosts extended near-zero magnetization regions and more complex multi-domain textures over several microns (Figs.~\ref{Fig2}b--j,~\ref{Fig4}a--m,~\ref{Fig_em_2}). A natural interpretation is that in the zero-field states superconductivity occupies relatively small regions near MDWs, so only a small fraction of each segment is superconducting, whereas in the coercive window a larger fraction of the segment lies in magnetically textured regions where the effective Zeeman field is reduced, with possibly both DWS and MDAS contributing to the observed superconductivity.

From the scanning SQUID images, our spatial resolution of $\approx1$~$\mu$m prevents a direct determination of the MDW width $d_\mathrm{w}$. Thin-film studies estimate $d_\mathrm{w} \approx 50$--$71$ nm in EuS~\cite{Houzet2006Dec,Aikebaier2019Mar}, but these values pertain to thicker, planar films. In our ultrathin, curved EuS shell, geometry and reduced thickness can substantially modify $d_\mathrm{w}$. Even if $d_\mathrm{w} < \xi \approx 200$ nm for full Al coatings in such nanowires~\cite{Vaitiekenas2020Mar,Vaitiekenas2020Feb}, theory predicts that robust DWS can still occur for $d_\mathrm{w} \lesssim \xi$ in ferromagnet-superconductor heterostructures with exchange interactions~\cite{Houzet2006Dec}.
Alternative explanations such as a Little-Parks effect are unlikely. Unlike the zero-field-centered, field-periodic $\mathrm{d}V/\mathrm{d}I$ oscillations reported in InAs/Al full-shell nanowires~\cite{Vaitiekenas2020Mar,Vaitiekenas2020Feb, Valentini2021Jul}, our data show no superconductivity at $H_\mathrm{a}=0$ and a non-periodic, hysteretic field response tied to the EuS coercivity. The finite, hysteretic field window in which superconductivity appears is incompatible with the period set by the Al shell’s effective area $A_\mathrm{eff}$, $\Delta H \simeq \Phi_0/A_\mathrm{eff}$, which in our geometry is at least on the order of tens of mT. Although the EuS can shift the oscillation center away from zero field, it cannot account for the lack of periodicity or the mismatch with $\Phi_0/A_\mathrm{eff}$.

\section{Conclusion and Outlook} 
In conclusion, we have shown that superconductivity in the Al shell of full-shell InAs/EuS/Al nanowires is tied to the magnetic texture of the EuS layer. From scanning SQUID magnetometry and differential resistance measurements, we find that superconducting signatures appear only when the EuS is in a multi-domain state, after zero-field cooling and near coercivity, and are absent in the saturated single-domain state. Micromagnetic simulations of the EuS hysteresis further support this picture. Comparing different nanowires, transport segments, and field angles, we conclude that the observed behavior is compatible with two microscopic scenarios, domain wall superconductivity (DWS) near MDWs and multi-domain-averaged superconductivity (MDAS), which our present measurements cannot distinguish between.

Our experimental geometry opens promising avenues for future device architectures utilizing the spatial control of MDWs and related DWS. 
For example, recent theoretical advances suggest that a mobile racetrack MDW embedded in a Josephson weak link can control and reverse the superconducting diode effect, enabling a Josephson-transistor response that serves as a fast, cryogenic, low-dissipation readout of racetrack states~\cite{HessNov2023}. 
InAs/Al nanowire platforms have already realized superconductor-quantum-dot junctions and Josephson weak links~\cite{Saldana2020Feb,Zellekens2020Apr}, whereas we show spatially controllable MDWs in EuS, together pointing to a possible experimental route to realize the proposal.
Furthermore, the ability to locally manipulate the superconducting phase enables the engineering of phase-biased junctions without external magnetic fields. 
For Andreev spin qubit recently realized in InAs/Al nanowire Josephson weak links, this can offer on-chip, reconfigurable phase control that supports spin-selective readout and may improve coherence and device scalability \cite{Hayes2021SpinQubitNW}.

\begin{acknowledgments}
The authors acknowledge experimental assistance from Daria Beznasiuk and useful discussions with Charles M. Marcus and Yusuke Iguchi. The project received funding from the European Union’s Horizon 2020 research and innovation programme under the Marie Sklodowska-Curie grant agreement No.~832645 (SpinScreen) and No. 722176 (INDEED), the ERC starting Grant No.~716655, Carlsberg Foundation, Villum Foundation project No.~25310, the Danish National Research Foundation (DNRF-101), the Sino-Danish Center for Education and Research and Microsoft.
\end{acknowledgments}

\bibliography{Ref}

\begin{table*}[tp]
\centering
\caption{MBE-grown hybrid nanowires used in the experiments. Nominal incident angle ($\theta$), thickness ($d$), and sequence of EuS and Al in-situ deposition after nanowire growth, and the devices in which the nanowires were used. To prepare the nanowire used in device F, both EuS and Al were deposited twice, bringing their total thicknesses to 10 nm each.}
\label{Tab1}
\begin{tabular}{|l|c|c|c|c|c|c|c|c|c|}
\hline
\multirow{2}{*}{Device segment} & \multirow{2}{*}{MBE Growth ID}
  & \multicolumn{2}{c|}{1. EuS}
  & \multicolumn{2}{c|}{2. EuS}
  & \multicolumn{2}{c|}{3. Al}
  & \multicolumn{2}{c|}{4. Al} \\ \cline{3-10}
 & & $\theta$ (deg) & $d$ (nm)
   & $\theta$ (deg) & $d$ (nm)
   & $\theta$ (deg) & $d$ (nm)
   & $\theta$ (deg) & $d$ (nm) \\ \hline
A, B, C & QDev951 & 75 & 3 & -- & -- & 82 & 4 & -- & -- \\ \hline
D, E    & QDev923 & 67 & 3 & -- & -- & 72 & 8 & -- & -- \\ \hline
F       & QDev972 & 75 & 3 & 82 & 7 & 80 & 5 & 85 & 5 \\ \hline
\end{tabular}
\end{table*}

\begin{figure*} [tp]
\centering
\includegraphics[width=0.9\textwidth]{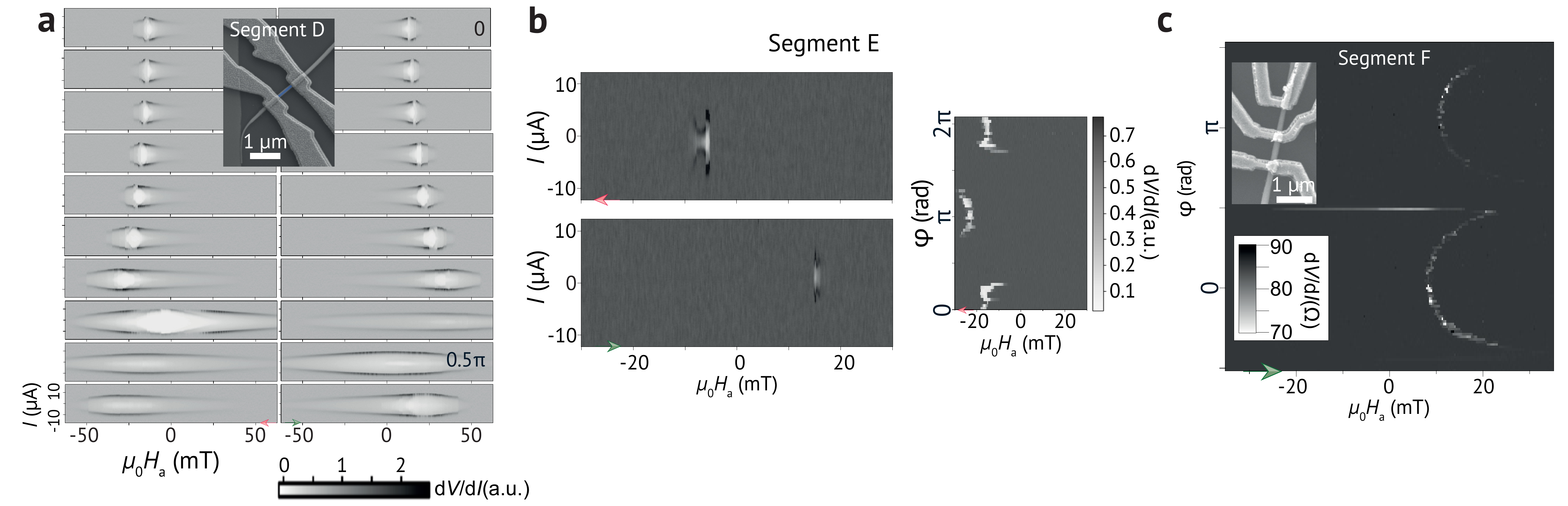}
\caption{{\bf{Additional transport measurements on InAs/EuS/Al full-shell nanowire devices while sweeping an applied magnetic field and field angle.}} 
}
\label{Fig_em_1}
\end{figure*}

\onecolumngrid
\vspace*{2ex}
\begin{center}
\textbf{\large End Matter}
\end{center}
\vspace*{2ex}
\twocolumngrid

{\bf{Nanowire growth and device fabrication}} The nanowires were grown from etched pits in InAs substrates in order to facilitate shadow deposition of the EuS/Al shells~\cite{Carrad2020Apr,Khan2020Mar}. After MBE growth of the core InAs nanowires, deposition of EuS and Al was done by performing multiple electron beam evaporations without breaking the vacuum~\cite{Liu2020Feb}. By using different incident angles, the shells coated only the upper section of the nanowires. EuS deposition was carried out at 180 °C as measured with a thermo-couple sensor, while Al was subsequently deposited after cooling the sample holder below -20 °C. For further details, see SM. 
Table~\ref{Tab1} summarizes the growth parameters for the measured nanowire devices.

{\bf{Transport measurements from additional nanowire devices}} In Fig.~\ref{Fig_em_1}, we show data from three segments in two devices with thicker Al coatings. As $\phi$ is swept away from zero, all three segments show a shift in both the field value and the field range over which the superconducting phase appears, consistent with segment B in the main text. For segments E and F, from two different devices, no superconductivity is observed in zero-bias $\text{d}V/\text{d}I$ over a range of angles near $\phi = 0.5\pi$. SQUID images at these larger field angles show that, at the highest applied fields, the EuS remains in a multi-domain state in which the domains tend to align along the applied field and therefore lie predominantly transverse to the nanowire axis, with sub-micron to micron scale domains. As the field is reduced along the hysteresis loop, the domains gradually rotate back toward the nanowire axis while remaining multi-domain over micron scales. In this geometry, a $\approx 700$~nm transport segment along the nanowire may lie entirely within a single domain throughout the hysteresis loop, in which case it may never sample a multi-domain configuration and superconductivity can be absent. Alternatively, it may overlap a region where MDWs and/or multiple domains intersect the segment over a broad range of fields, in which case superconductivity can persist across nearly the entire field sweep. Taken together with the data presented in the main text, these measurements suggest that superconducting signatures vary substantially between transport devices and segments, both in whether they appear at all and in the extent and shape of the superconducting region as a function of field-sweep direction, field angle, and bias current.

{\bf{Extracting the hysteresis loop from scanning SQUID magnetometry images}}
The SQUID measures only the out-of-plane flux. If the magnetization is not strictly confined to a 2D plane parallel to the SQUID scan plane, knowing the out-of-plane component of the flux alone cannot uniquely reconstruct the magnetization pattern. In our hexagonal cross-section nanowires the magnetization lies in the plane of each facet; because the facets are tilted relative to the scan plane and sit at different distances from the SQUID, the measured flux contains geometry-dependent projections and cannot be uniquely inverted. A fully quantitative reconstruction would also require the exact nanowire dimensions. For these reasons, we report $\tilde m_\text{z}(H_\mathrm{a})$ as a normalized measure of the axial magnetic moment rather than an absolute number.

\begin{figure*} [tp]
\centering
\includegraphics[width=\textwidth]{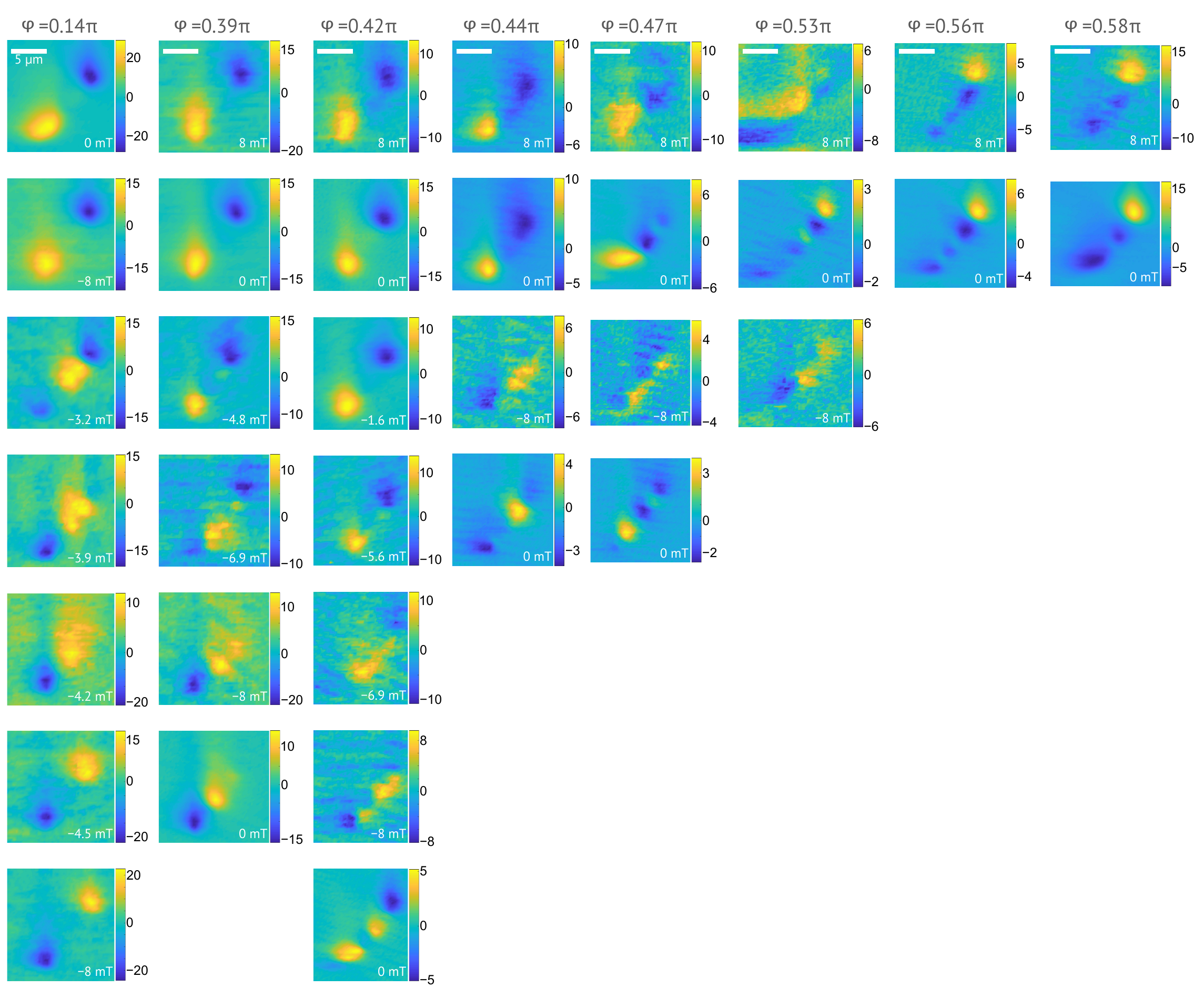}
\caption{{\bf{SQUID magnetometry images of EuS domain formation in NW2 as a function of field angle.}} SQUID magnetometry images of the EuS shell in NW2 at eight field angles, $\phi = 0.14\pi$, $0.39\pi$, $0.42\pi$, $0.44\pi$, $0.47\pi$, $0.53\pi$, $0.56\pi$, and $0.58\pi$. For each $\phi$, $H_\mathrm{a}$ is swept from $0 \to +8$~mT, $+8 \to -8$~mT, and $-8 \to 0$~mT, and images at selected fields along these sweeps are shown. All scans are taken at 4.2 K. 
}
\label{Fig_em_2}
\end{figure*}

To obtain a hysteresis loop from the SQUID images, we treated the nanowire as a 1D Ising chain that magnetizes only along its axis $\hat{z}$. For each applied field $H_\mathrm{a}$, we divided the nanowire into axial domains from the measured flux polarity, measure each domain’s length $L_i(H_\mathrm{a})$, and compute
\begin{equation}
\tilde{m}_\text{z}(H_\mathrm{a}) \;=\; \frac{\sum_i s_i\, L_i(H_\mathrm{a})}{L_\text{tot}} \,, \qquad s_i = \pm 1.
\end{equation}
Here $L_{tot}$ is the total length of the single domain at saturation, i.e. the length of the nanowire.

{\bf{Additional angle dependent SQUID magnetometry images}} Fig.~\ref{Fig_em_2} summarizes how the EuS domain structure in NW2 evolves with field angle and provides additional SQUID magnetometry images complementing Fig.~\ref{Fig3}.
At eight representative field angles between $\phi = 0.14\pi$ and $0.58\pi$, the images illustrate how the domain formation changes with field angle. 
For each angle, $H_\text{a}$ is swept through a full hysteresis loop, 0~mT to $+8$~mT, to $-8$~mT, and back to 0~mT.
All scans are taken at 4.2 K.

\end{document}